# Infrastructure Disparities in Northern Malaysia

*Siti Hadijah Che Mat\*, School of Economic, Finance and Banking, Universiti Utara Malaysia Kedah, Malaysia.*

*E-mail: hadijah@uum.edu.my*

*Mukaramah Harun, School of Economic, Finance and Banking, Universiti Utara Malaysia Kedah, Malaysia.*

*Wan Roshidah Fadzim, School of Economic, Finance and Banking, Universiti Utara Malaysia Kedah, Malaysia.*

*Mohd Saifoul Zamzuri Noor Norzita Jamil, School of Economic, Finance and Banking, Universiti Utara Malaysia Kedah, Malaysia.*

*Nor'Aznin Abu Bakar, School of Economic, Finance and Banking, Universiti Utara Malaysia Kedah, Malaysia.*

**Abstract**--- This study examines the disparities of infrastructure in four states in Northern Peninsular Malaysia. This study used a primer data which is collected by using a face to face interview with a structure questionnaire on head of household at Kedah, Perlis, Penang and Perak. The list of respondents is provided by the Department of Statistics of Malaysia (DOS). The Department of Statistics of Malaysia (DOS) uses the population observation in 2010 to determine the respondents' home position. In order to see the disparity between the state, DOS has issued 400 respondents for urban and rural areas. The results of the study found in terms of transport infrastructure, most of the states show an equal percentage in term of the distance of their houses to the highway except for Perlis with 50% respondents' house are more than 20km to the highway. The majority of respondents have a paved road to their houses, meanwhile it is noticed that in Kedah and Perak, 1.6% and 0.7% of the respondents have a red dirt road to their houses.

**Keyword**s--- Infrastructure, Disparities, Northern Malaysia.

## I. Introduction

Malaysia has achieved the independence for more than 60 years. However, the economic situation in Malaysia can still be distinguished by state and strata. The state in the western region of Malaysia is developing rapidly compared to the state in the eastern part of the country. Similarly to the poverty and inequality of income, the poverty rates in each state in Peninsular Malaysia vary between rural and urban area. Therefore, under the Ninth Malaysian Plan, five economic corridors were initiated to address imbalances throughout the country with clear objectives such as to accelerate the movements towards high value, knowledge-driven economic activities and high-income economy, to promote balanced development, etc. The master plan target for new jobs creation is 1.57 million and investment is RM178 billion. Within five years (2007-2012), the achievement in terms of job creation was 0.42 million and investment was RM29.70 billion.

M. Sharif, *et.al* (2010), in their study stated that the development gaps between states are reflected in the attractiveness of new manufacturing investment in certain states where the direction of investment was skewed towards the more developed states due to the availability of adequate and good infrastructure. Infrastructure is the main driving force of development for any country. The importance of physical infrastructure in economic development, employment and reducing disparity within the country/region has been recognized by many researchers.

Availability of adequate infrastructure facilities, especially the physical infrastructure is an important pre-condition for sustainable economic and social development. The development of a region is disturbed by the low quality and limited accessibility of infrastructure. It is also understood that the investment in infrastructure projects involves huge capital, long gestation periods, high incremental capital-output ratio, high risk and low rate of return on investment. All of these factors make private sector entry difficult. Poor infrastructure not only leads to underutilise the region's economic resources but also fail to market, the domestic produces at the fullest level in developing countries.

Generally, we can divide infrastructure into two categories, which is economic and social infrastructure. Economic infrastructure includes transport, communication, irrigation, energy, banking, etc., whereas social infrastructure include sectors such as health, education, housing, etc.

Regional disparities in economic development can be explained in term of different levels of infrastructural services available to people in different regions. Improvement in infrastructural services is crucial to enhance efficiency of the production process and to raise productivity of any economic entity. Development economists have argued that physical infrastructure being a prerequisite for industrialization and successful economic development, (Murphy, Shleifer, and Vishny 1989; Sawada, 2015).

Within the context of Malaysia, the issue of infrastructure development and regional disparity need to be examined further as infrastructure development may have a disproportionate positive impact on the income and welfare of the people. Conceptually, infrastructure helps poorer individuals and underdeveloped areas to get connected to core economic activities, thus allowing them to access additional product opportunities (Estache, 2003). Likewise, infrastructure development in the poorer regions reduces production and transaction costs (Gannon and Liu, 1997). Estache and Fay (1995) in their study on Argentina and Brazil find that enhanced access to roads and sanitation has been a key determinant of income convergence for the poorest regions. Along the same lines, infrastructure access can raise the value of the assets of the poor. Improvements in communication and road services imply capital gains for these poor farmers (Jacoby, 2000). Infrastructure development can also have a disproportionate impact on the human capital of the poor, and hence on their job opportunities and income prospects. This refers not only to education, but most importantly to health. However, the new economic geography model postulates that variations in the availability and quality of infrastructure across space will result in different economic agents' behaviour depending on their location. Moreover, they will also crucially influence agents' location decisions, such as migration, establishment of new firms, etc.

Achieving balance regional development still remains as one of the prime objectives of the Malaysia development plan. There exists a regional disparity amongst states in Malaysia and one of the reasons is because of inadequate infrastructure. It is obvious that in those areas where there is insufficient economic and social infrastructure, the regions or areas usually considered poor and those poorer regions cannot contribute properly in the process of economic growth and development. This article aims to analyze Inter-State and Intra-State Infrastructural Disparities in four states in the northern region of the Peninsular Malaysia.

## II. Methodology

This study uses primary data collected through surveys on household heads in urban and rural areas in four states of the Malaysia peninsula. The list of respondents is provided by the Department of Statistics of Malaysia (DOS). The Department of Statistics of Malaysia (DOS) uses the population observation in 2010 to determine the respondents' home position.

For the purpose of seeing the disparity between the state, DOS has issued a list of respondents for the four states, namely Kedah, Perlis, Pulau Pinang and Perak. For each state, DOS has set the number of respondents to be interviewed in urban and rural areas. Data for four main sectors of infrastructure services, namely, transport, ICT, electricity, water and sanitation are used in this study. Few sub-indicators are included in each sector to ensure better representation of different aspects of infrastructure provision and availability.

After knowing the number of samples then the next process is to do the field work, i.e. a total of 400 questionnaires were distributed and all were quoted and analyzed as data collection was conducted face-to-face between the enumerator and the respondents. But after the data cleaning process, only 392 can be analyzed. This is because the rest there is some less information. The data is analyzed and presented in the form of tables, frequencies, and percentages.

## III. Finding

*Demographic Characteristics*

Out of 400 respondents, only 392 were analysed. Table 1 presents the sample characteristics of the study. As can be seen from Table 1, the majority of respondents are between age 41 – 50 years old (23.7%). They were 61.5% male respondents and 38.5% female respondents.

The majority of respondents (80.9%) are married. 9.7% respondents are single, 8.7% respondents have divorced/widow/widower and 0.8% respondents were separated.

Meanwhile, the sample is dominated by Malay (75.0%). Second highest respondents are Chinese (16.3%), followed by Indian (8.2%) and other ethnic 0.6%.

Table 1: Respondents Characteristics

| Characteristics | | Frequency | % |
|---|---|---|---|
| Age | 20 and below | 9 | 2.3 |
| | 21-30 | 31 | 7.9 |
| | 31-40 | 88 | 22.4 |
| | 41-50 | 93 | 23.7 |
| | 51-60 | 92 | 23.5 |
| | 61-70 | 57 | 14.5 |
| | 71 and above | 22 | 5.6 |
| Gender | Male | 241 | 61.5 |
| | Female | 151 | 38.5 |
| Ethnicity | Malay | 294 | 75.0 |
| | Chinese | 64 | 16.3 |
| | India | 32 | 8.2 |
| | Others | 2 | 0.6 |
| Marital status | Single | 38 | 9.7 |
| | Married | 317 | 80.9 |
| | Widow/Widower/Divorced | 34 | 8.7 |
| | Separated | 3 | 0.8 |
| Level of Education | No formal education | 11 | 2.8 |
| | Primary school (standard 1-6) | 91 | 23.2 |
| | Secondary school (form 1-3) | 51 | 13.0 |
| | Secondary school (vocational/Technic) | 3 | 0.8 |
| | Secondary school (form 4-5) | 134 | 34.2 |
| | Secondary school (lower 6-higher 6/Matriculation) | 18 | 4.6 |
| | | 43 | 11.0 |
| | Politechnic/College/University | 41 | 10.5 |
| Number of children (included adopted children / stepchildren) | 0 | 40 | 10.2 |
| | 1 | 43 | 11.0 |
| | 2-3 | 141 | 36.0 |
| | 4-6 | 111 | 35.8 |
| | 7-9 | 50 | 6.4 |
| | >=10 | 3 | 0.9 |
| Number of dependents | 0 | 65 | 16.6 |
| | 1-2 | 143 | 36.4 |
| | 3-4 | 131 | 33.4 |
| | 4-6 | 95 | 24.2 |
| | >=7 | 15 | 3.9 |
| Main income (RM) | Below 1000 | 139 | 35.5 |
| | 1000-2000 | 145 | 37.0 |
| | 2001-3000 | 56 | 14.3 |
| | 3001-4000 | 21 | 5.4 |
| | 4001-5000 | 17 | 4.3 |
| | >5000 | 14 | 3.6 |

**Source**: *Authors' computations from questionnaire survey*

In term of level of education among respondent, the majority of them have education until secondary school (form 4-5) (34.2%). The second highest percentage of respondents, have education until primary school (standard 1-6) (23.2%). They were also respondents who did not have any formal education with 2.8% while they were 10.5% respondents who received the education from the university. From the table also shows respondent's main income from their main job. The majority of respondents earn between RM1000-RM2000 (37.0%) monthly. The lowest of income in the category is below RM1000 and they were 35.5% respondents in this category.

*Inter-State Infrastructural Disparities*

Table 2 provides a comparison percentage between all states. In terms of transport infrastructure, most of the states show an equal percentage in term of the distance of their houses to the highway except for Perlis with 50% respondents' house are more than 20km to the highway. The majority of respondents have a paved road to their houses, meanwhile it is noticed that in Kedah and Perak, 1.6% and 0.7% of the respondents have a red dirt road to their houses.

Most prior studies highlight the significance of infrastructure development in the process of socioeconomic development. Indeed, the theoretical evaluation of the impact of infrastructure development on economic development is closely connected to growth theory. For example, the study of Arrow and Kurz (1970) includes infrastructure into the growth theory literature. Infrastructure measured by the public capital has been considered as an additional important input in the production function.

From a policy perspective, this study suggests that infrastructure development contributes positively to economic growth in the northern states of Malaysia 'Northern Corridor Economic Region (NCER)'. In this context, the NCER's aggressive investment on infrastructure is justified to sustain growth. The NCER has taken a necessary action in looking at the infrastructure needs of the region supporting by the federal government funds allocated for infrastructural improvements. Improving the regional and international connectivity, the infrastructure investment of RM20bn was channeled by the Malaysian government into the NCER and this crucial investments involve the Ipoh-Padang Besar double-tracking project to upgrade rail capacity and the second Penang Bridge.

With better access to roads and other means of public transportation, i.e. railways, workers can get to their job more easily, thus spending less time commuting from home or moving across different work locations. This would tend to reduce traffic-related stress, which can be detrimental to concentrate on the job. In addition, with greater access to electricity and telecommunications, the worker can perform numerous tasks more promptly as well as additional tasks away from the office, therefore leads to higher productivity which will enhance growth.

The forecasting demand for infrastructure shows an increasing trend which indicate that the policy makers need to formulate a policy that can encourage investment in these sectors. The authority must ensure that the water supply is sufficient according to demand so that the region will not suffer from water shortages in the future. Investing in these sectors can further boost the economic growth by enhancing the level of human capital. As stated in several studies, access to safe water and sanitation helps to improve health, particularly among children. Besides that, access to clean water and sanitation infrastructure helps to reduce infant mortality.

Moreover, as the country is moving towards high income country, the demand for ICT i.e. broadband/internet will definitely increase.

However, providing broadband in rural areas poses significant economic and technical challenges. Costs in areas of low population density are higher and, unlike other ICTs, the provision of broadband has technical constraints by which available speeds diminish with increasing distance from a central location. The speedy development of the broadband market has therefore focused primarily on urban centres leaving the majority of people in rural areas unable to access network services. As public and private services are increasingly provided online, the inability for some parts of the population to get access to broadband becomes more of a public policy problem. Once broadband usage reaches a critical mass it will come to be considered vital for all, if balanced development is to be achieved without discrimination based on geographical location. Therefore the governments need to consider a more vigorous approach to ensuring broadband is available throughout the areas.

Finding related to ICT show, the majority of respondents have an equal percentage of having the internet, fixed telephone line, computer/laptop except in Kedah where only 29.7% respondents have a computer/laptop. The majority of respondents stated almost equal percentage about the internet problem that occurs for each problem.

Further, all respondents have electricity in their house. Respondents also seem to have an almost equal percentage of problem regarding electricity supply except in Perak, where only 66.7% have a problem of low voltage. Becoming a high income country in year 2020, it is also expected that the demand for energy/electricity will increase. Securing the supply of electricity will be crucially important to meeting the increasing growth in electricity consumption and to sustain economic growth.

The authority must ensure that the transmission and distribution of electricity supply is adequate and reliable for the economic activities to perform efficiently. In addition, an increase in energy capacity, the building-up of energy supply infrastructure and contingency planning in the event of a technical disruption will be essential, notwithstanding the drive to use more renewable sources of energy.

In terms of water and sanitation, majority and family used water pipe at home as the source of drinking water, for cook and other purpose. From the above analysis, it is noticed that most of the states behave almost equally, therefore it is not much infrastructural disparity in the four states.

Table 2 also shows source of water used by the family. From the table above, the majority of respondents used water pipe at home.

They were 100% of respondents used water pipe at home in Perlis and Perak, 92.2% of respondents in Kedah, 91.3% respondents in Pulau Pinang. They were only some respondents who used other than water pipe at home as a source of drinking water, such as in Kedah, they were 7% respondents used well/spring- and in Pulau Pinang,1.0% used bottled/water filtered.

Table 2: Inter-State Infrastructural Disparities

| INDICATORS | KEDAH (n=128) | PERLIS(n=16) | P.PINANG(n=104) | PERAK(n=144) |
|---|---|---|---|---|
| **TRANSPORTATION** | | | | |
| **Distance from house to highway** | | | | |
| 1-5km | 48 (37.5) | 5 (31.2) | 35 (33.70) | 41(38.5) |
| 6-10km | 30 (23.4) | 3 (18.8) | 39 (37.5) | 36 (25) |
| 11-15km | 17 (13.3) | 0 | 6 (5.8) | 23 (16.0) |
| 16-20km | 1(0.8) | 0 | 14 (13.5) | 5 (3.5) |
| More than 20km | 37(28.9) | 8 (50) | 10 (9.6) | 39 (27.1) |
| **Type of road to respondent's house** | | | | |
| Paved road | 110 (85.9) | 16 (100.0) | 99 (95.2) | 143 (99.3) |
| Gravel road | 16 (12.5) | 0 (0.0) | 5 (4.8) | 0 (0.0) |
| Red Dirt Road | 2 (1.6) | 0 (0.0) | 0 (0.0) | 1 (0.7) |
| **ICT/COMMUNICATION** | | | | |
| **Item related to ICT** | | | | |
| Internet | 29 (22.7) | 10 (62.5) | 69 (66.3) | 94 (65.3) |
| Fixed telephone Line | 28 (21.9) | 14 (87.5) | 50 (48.1) | 53 (36.8) |
| Laptop/Computer | 38 (29.7) | 8 (50.0) | 54 (51.9) | 82 (56.9) |
| **Internet Problem** | | | | |
| Limited coverage | 5 (22.7) | 1 (25.0) | 7 (21.2) | 9 (23.1) |
| Expensive Charge | 1 (4.5) | 1 (25.0) | 9 (27.3) | 9 (23.1) |
| Unstable coverage | 16 (72.7) | 2 (50.0) | 15 (45.5) | 19 (48.7) |
| Others | 4 (3.1) | 0 (0.0) | 2 (6.1) | 2 (5.1) |
| **ELECTRICITY** | | | | |
| Electricity | 128 100.0) | 16 (100.0) | 104 (100.0) | 144 (100.0) |
| **Problem of electricity Supply** | | | | |
| Electric Rationing | 2 (15.4) | 0 (0.0) | 2 (28.6) | 1 (8.3) |
| Low voltage | 6 (46.2) | 3 (100.0) | 1 (14.3) | 8 (66.7) |
| Variable Voltage | 5 (38.5) | 0 (0.0) | 1 (14.3) | 2 (16.7) |
| Others | 0 (0.0) | 0 (0.0) | 3 (42.9) | 1 (8.3) |
| **WATER & SANITATION** | | | | |
| **Source of water** | | | | |
| Water pipe at home | 118 (92.2) | 16 (100.0) | 95 (91.3) | 143 (99.3) |
| Neighbor's/public/well/spring | 9 (7.0) | 0 (0.0) | 0 (0.0) | 1 (0.7) |
| Bottled/Filtered Water | 0 (0.0) | 0 (0.0) | 9 (8.7) | 0 (0.0) |
| Others | 1 (0.8) | 0 (0.0) | 0 (0.0) | 0 (0.0) |

## IV. Conclusion

The public sector has to play a leading role through direct investment in various infrastructure services. The government also should encourage more participation of private sectors to invest through policies that tend to enhance incentives for private investors. Special tax rebates can be given to the existing private investors and to attract new investors. This strategy is very important for the survival and growth of the infrastructure sectors.Public sector also should take initiatives in the form of public-private partnerships (PPP) for the projects that have potentials, but the private sector is hesitate to invest due to higher risk. These PPP can open opportunities for new investment projects and attract new investors.

The significance of infrastructure in Malaysia has been encouraged as early as the Ninth Malaysia Plan (2006-2010) and bolstered with real initiatives and action plans in the Tenth Malaysia Plan (2011-2015) and Eleventh Malaysia Plan (2015-2020). The study of Riaz (1997) describes the pattern of telecommunications infrastructure development and services, specifically in the late-1980s in Malaysia. The development of telecommunications infrastructure plays a key role in the development of the Malaysian economy. Lee (2011) mentions that the value of the infrastructure sector to economic growth and development has long been documented and understood by academicians and policymakers. The aim of the study provided insights on how the infrastructure sector contributes to the Malaysian economic development.

The analysis on survey data collected, reveals that there is not much infrastructural disparity between states. The infrastructural disparities are observed in most of the infrastructure services for instance, in transportation facilities, rural areas show a bigger percentage in terms of distance from the respondent's house to the highway, as well as other services such as broadband and internet.